\pdfoutput=1

\documentclass[11pt]{article}

\usepackage[final]{acl}

\usepackage{times}
\usepackage{latexsym}

\usepackage{microtype}

\usepackage[T1]{fontenc}

\usepackage[utf8]{inputenc}

\usepackage{inconsolata}

\usepackage{graphicx}

\usepackage{amsmath}
\usepackage{amsfonts}
\usepackage{amssymb}
\usepackage{algorithm,algorithmic}
\usepackage{float}
\usepackage{color}
\usepackage{xcolor}
\usepackage{multirow}
\usepackage{bigdelim}

\title{Tethering Broken Themes: Aligning Neural Topic Models \\ with Labels and Authors}

\author{Mayank Nagda \and Phil Ostheimer \and Sophie Fellenz\\
  RPTU Kaiserslautern-Landau, Germany\\
  \texttt{surname@cs.uni-kl.de}}

\begin{document}
\maketitle
\begin{abstract}
Topic models are a popular approach for extracting semantic information from large document collections. However, recent studies suggest that the topics generated by these models often do not align well with human intentions. Although metadata such as labels and authorship information are available, it has not yet been effectively incorporated into neural topic models. To address this gap, we introduce FANToM, a novel method to align neural topic models with both labels and authorship information. FANToM allows for the inclusion of this metadata when available, producing interpretable topics and author distributions for each topic. Our approach demonstrates greater expressiveness than conventional topic models by learning the alignment between labels, topics, and authors. Experimental results show that FANToM improves existing models in terms of both topic quality and alignment. Additionally, it identifies author interests and similarities.
\end{abstract}

\section{Introduction}

Topic models are a family of generative models that help discover sets of words (called \textit{topics}) describing the semantics of a large document collection \cite{blei2003latent}. Topic models find applications in various fields, including healthcare \cite{rajendra2021visual}, political science \cite{grimmer2013text,karakkaparambil-james-etal-2024-evaluating}, psycholinguistics \cite{marcio24}, bioinformatics \cite{liu2016overview,doi:10.1142/9789811215636_0029}, among others. These models can be categorized into statistical models, such as latent Dirichlet allocation (LDA) \cite{blei2003latent}, or neural topic models (NTMs). NTMs are based on generative models such as variational autoencoders (VAEs) \cite{miao2016neural,wu2023effective,ijcai2024p710} or, more recently, Large Language Models (LLM) \cite{bianchi2020cross}. NTMs have been shown to learn topics with improved quality compared to statistical topic models \cite{hoyle2021automated}.

\begin{figure}[t]
    \centering
    \includegraphics[scale=0.85]{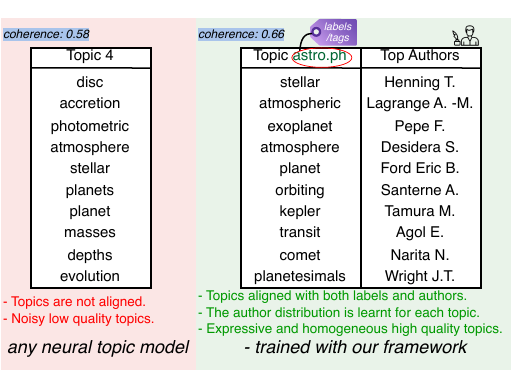}
    \caption{
    FANToM in action: A comparison of semantically closest topics learned by DVAE (left) and DVAE trained with FANToM (right) for alignment. Notably, FANToM not only accurately aligns the learned topic with the label (astrophysics) and authors but also improves the quality of the learned topic.
    }
    \label{fig:main-fig}
\end{figure}

However, recent studies reveal that current NTMs often fail to align well with human intentions and labeling  \cite{zhao2017metalda,doogan-buntine-2021-topic,hoyle2022neural}. For instance, in a scenario where the available labels are "Windows" and "MacOS," a misaligned model might merge them into a single "OS" topic, despite the intent to distinguish between these two labels. In other cases, where a general "OS" topic would suffice, the model might inappropriately split topics into specific operating systems.

Beyond labels, authorship information is also crucial for aligning topic models. Typically, an author focuses on a limited range of topics, and understanding these interests is fundamental for NLP and information retrieval tasks that involve large document collections \cite{zvi2004}. Modeling author interests enables us to answer key questions about document content, such as which subjects an author covers, which authors have similar writing styles, and which authors work on comparable topics \cite{tang2022contrastive, li2015author}. Although statistical models have been used to model author interests and link topics to authors \cite{zvi2004}, incorporating authorship information into NTMs remains a challenge.

\textbf{In this paper}, we present \textbf{FANToM}, a novel \textbf{F}ramework for \textbf{A}ligning \textbf{N}eural \textbf{To}pic \textbf{M}odels, which incorporates metadata such as labels and authorship information in existing NTMs. As shown in Figure~\ref{fig:main-fig}, FANToM aligns the learned topics with both the labels and the authors, enhancing the interpretability of the model. This approach not only establishes a connection between latent topics, labels, and authors, but also helps identify author interests based on the topics.

In summary, our contributions are as follows.
\begin{itemize}
    \item We introduce FANToM, a framework that aligns latent topics in NTMs with document labels and authors (Section~\ref{sec:method}).
\footnote{Code: \href{https://github.com/mayanknagda/fantom}{https://github.com/mayanknagda/fantom}} 
    \item We demonstrate through experiments (Section~\ref{sec:results}) that FANToM not only effectively aligns NTMs with labels and authors but also improves the quality of the learned topics, outperforming existing models.
    \item Our extensive experiments (Section~\ref{sec:discuss}) show how FANToM facilitates learning a shared embedding space for authors, words, and topics, and provides insight into author interests and similarities.
\end{itemize}

\section{Are Neural Topic Models Misaligned?}
\begin{figure}[ht!]
    \centering
    \includegraphics[scale=0.23]{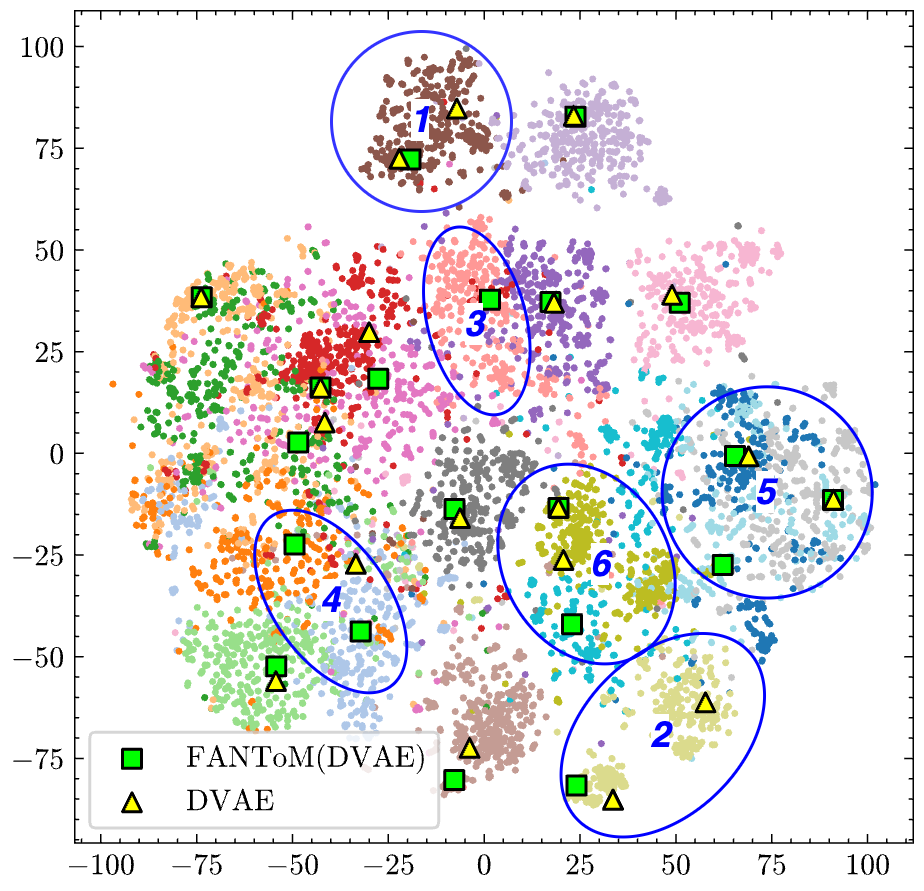}
    \caption{t-SNE projection of topic embeddings from the DVAE model (triangles) and its FANToM variant (squares), alongside document embeddings from the 20NG dataset, color-coded by labels. Ideally, topic embeddings should be positioned near the centroid of their corresponding document clusters. The circled regions highlight discrepancies where DVAE either overrepresents or underrepresents certain topics, while FANToM achieves a more balanced and accurate alignment with document labels, reinforcing its effectiveness in topic representation.}
    \label{fig:coverage}
\end{figure}
Recently, the alignment of discovered topics with human-determined labels has come under scrutiny \cite{zhao2017metalda, doogan-buntine-2021-topic, hoyle2022neural}. Despite decades of application in various domains, existing topic models struggle to align their generated topics with the intentions and expectations of human users. This discrepancy is particularly concerning, given that the primary objective of topic models is to uncover meaningful, interpretable patterns from text data. For instance, Hoyle et al. \shortcite{hoyle2022neural} highlighted issues of low topic purity and stability by utilizing human-assigned labels. Their findings indicate that numerous models fail to capture the nuanced distinctions that users make between different topics. 

To further investigate this issue, we analyze the document and topic embeddings generated by the baseline model, Dirichlet-VAE (DVAE) \cite{burkhardt2019decoupling}, and its FANToM variant. We obtain the document and topic embeddings by using SBERT \cite{reimers-gurevych-2019-sentence} which allocates the embeddings based on the content of the text. Ideally, topic embeddings should be positioned near the centroids of document embeddings, signaling high cluster purity and a strong correlation between topics and their respective documents. In Figure~\ref{fig:coverage}, using t-SNE \cite{van2008visualizing} projection we visualize these embeddings and pinpoint six problematic regions that reveal significant concerns regarding topic representation.

In regions 1, 2, and 6, we observe that the DVAE model tends to overrepresent certain clusters, leading to the generation of multiple topics for the same document label. This phenomenon not only diminishes topic diversity but also clouds the interpretability of the model's outputs. For example, when multiple topics are assigned to a single label, it becomes challenging for users to discern the unique contributions of each topic, ultimately undermining the model's utility.

Conversely, in regions 3, 4, and 5, the DVAE model demonstrates an underrepresentation of other clusters, resulting in the complete omission of some niche topics. This imbalance is detrimental as it indicates that the model is not adequately capturing the breadth of the data it is trained on. Such gaps in representation can lead to significant blind spots in the insights generated by the model, leaving users without a comprehensive understanding of the underlying themes present in the corpus.

\textbf{These observations lead to our main hypothesis.}
We hypothesize that the misalignment observed in neural topic models, is largely a consequence of the unconstrained nature of the latent topic distribution. In the absence of constraints, the model is free to allocate probability mass as it sees fit, which can result in overrepresentation of certain labels while simultaneously leading to the underrepresentation of others. This inherent instability can manifest itself in two problematic ways: the emergence of redundant topics that offer little new information and the failure to recognize and represent niche topics that are critical for a holistic understanding of the data.

By integrating metadata such as document labels and authorship, our proposed model, FANToM, imposes necessary constraints on the topic distribution. These constraints serve to enhance both the stability and alignment of the model, thereby improving its overall performance.

\section{Related Work}
In this section, we present related work on topic models and its alignment.
\subsection{Topic Models}

Topic models based on Latent Dirichlet Allocation (LDA) \cite{blei2003latent} were originally trained using variational inference or Gibbs sampling \cite{griffiths2004integrating}. Miao et al. \shortcite{miao2016neural} introduced neural topic models (NTMs) based on Variational Autoencoders (VAEs) with a Gaussian prior on latent topic variables, which were later extended with a Dirichlet prior \cite{srivastava2017autoencoding, burkhardt2019decoupling}. Other NTMs based on Generative Adversarial Networks (GANs) \cite{wang-etal-2020-neural-topic} and word embeddings (ETMs) \cite{dieng2020topic,wu2023effective} have also been proposed. NTMs generally outperform statistical models in terms of topic quality \cite{srivastava2017autoencoding} due to their more flexible generative distributions. Furthermore, NTMs are compatible with advances in deep learning, such as Large Language Models (LLMs) \cite{bianchi2020cross} and word embeddings \cite{dieng2020topic, wu2023effective}. Recent LLM-based topic models cluster document embeddings from LLMs using simple clustering methods, such as k-means \cite{grootendorst_bertopic_2022} or Gaussian mixture models \cite{sia-etal-2020-tired}. However, some researchers do not consider these clustering methods as topic models because they do not produce document-topic distributions \cite{wu2023effective}. An example of an LLM-based topic model is CTM \cite{bianchi2020cross}, which uses a VAE for clustering document embeddings. Recent state-of-the-art ETMs, such as ECRTM \cite{wu2023effective}, cluster word embeddings with topic embeddings as centers using soft assignment. Our framework allows for the first time the alignment of all variants of NTMs based on VAEs, and we compare it to these models in our experiments.

\subsection{Alignment of Topic Models}

Topic model alignment involves ensuring that latent topics are aligned with metadata such as labels or authorship information \cite{rahimi2023antm, chuang2013topic, abels2021focusing}. This is typically achieved through supervision using metadata \cite{ramage-etal-2009-labeled, zvi2004}. Topic model supervision can be divided into methods to improve interpretability with metadata and methods with the goal of improving classification \cite{Burkhardt:2018-1,Burkhardt:2019-1,Burkhardt:2019-3}. However, with advancements in LLMs, the use of topic models for classification has become less relevant. We focus on supervising models to improve alignment and interpretability.

\paragraph{Alignment of Statistical Topic Models.} Labeled LDA learns one topic per label \cite{ramage-etal-2009-labeled}, while the Author-Topic Model (ATM) learns topic distributions for each author \cite{steyvers2004probabilistic, zvi2004}, enabling the computation of author similarities. Variants include the Author-Conference-Topic (ACT) model \cite{tang2008topic} and the Author-Recipient-Topic (ART) model \cite{mccallum2005topic}, which adds recipient-based analysis for emails. We use Labeled LDA and ATM as baselines in our experiments.

\begin{figure*}[tb]
    \centering
    \includegraphics[scale=0.92]{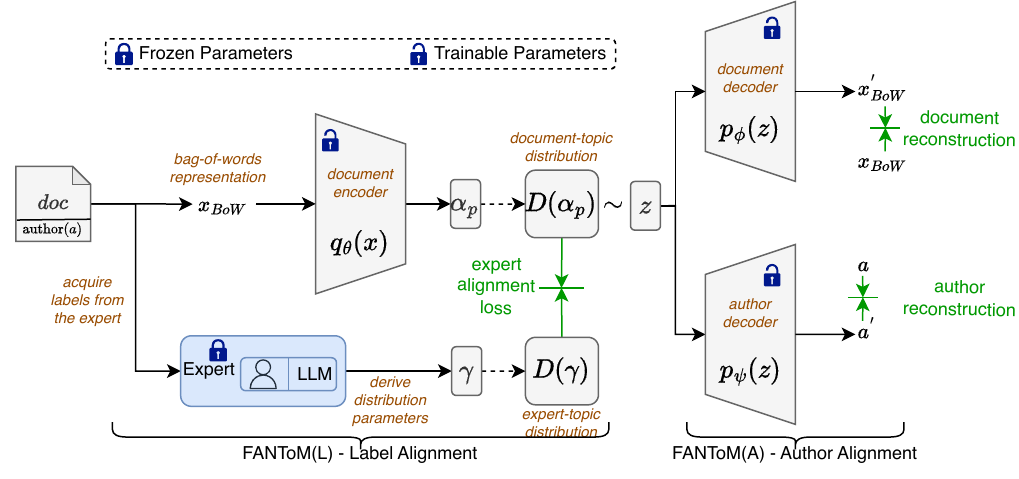}
    \caption{
    Illustration of FANToM: The framework aligns labels and authorship information with topics. It incorporates expert-assigned labels to establish a prior distribution parameterized by $\gamma$, which is then aligned with the posterior. For authorship, a separate decoder is used to learn the multinomial distribution over authors, ensuring a structured representation of author-topic relationships. Overall, FANToM ensures a structured and interpretable alignment between topics, labels, and authors.}
    \label{fig:architecture}
\end{figure*}

\paragraph{Alignment of Neural Topic Models.} Neural topic models (NTMs) have seen increasing development \cite{miao2016neural, srivastava2017autoencoding, burkhardt2019decoupling}; however, the number of supervised NTMs remains limited. SCHOLAR \cite{card-etal-2018-neural} uses metadata as input labels and constructs a classifier network from the latent vector to predict labels, generating topics relevant for classification. Rahimi et al. \shortcite{rahimi2023antm} proposed an aligned neural topic model for dynamic, evolving topics, which contrasts with our focus on static topics. TAM \cite{wang2020neural} trains an RNN classifier jointly with an NTM to predict labels from word sequences. Other models emphasize classification over alignment \cite{bai2018neural, korshunova2019discriminative}. Generally, NTMs do not enforce topic-label alignment, and aligned variants of supervised NTMs have not been thoroughly explored. Studies reveal that current NTMs often fail to align well with human-defined labels \cite{zhao2017metalda, doogan-buntine-2021-topic, hoyle2022neural}. In our experiments, we compare FANToM to supervised NTMs such as SCHOLAR and TAM.

\section{Methodology}
We begin by discussing the background in Sec.~\ref{sec:background} and present our proposed method in Sec.~\ref{sec:method}.
\subsection{Background}
\label{sec:background}
VAE-based topic models use an encoder-decoder architecture. Let $\{x_i\}_N$ represent the observed input documents in the Bag-of-Words (BoW) format, where $x_i \in \mathbb{N}^V$ and \( V \) is the vocabulary size. The encoder, parameterized by \( \theta \), maps the input to a latent vector \( z \), while the decoder, parameterized by \( \phi \), reconstructs the documents \cite{burkhardt2019decoupling, srivastava2017autoencoding}.

The objective is to learn the parameters \( \theta \) and \( \phi \) by minimizing the $\beta$-VAE loss:
\begin{equation}
\begin{aligned}
\mathcal{L}_{\text{vae}}\left(\theta, \phi ; x\right)&=-\mathbb{E}_{q_{\theta}\left(z \mid x\right)}\left[\log p_{\phi}\left(x \mid z\right)\right] \\
&+ \beta D_{K L}\left[q_{\theta}\left(z \mid x\right) \| p_{\alpha}\left(z\right)\right],
\end{aligned}
\label{eq:b-vae}
\end{equation}
where the first term is the reconstruction loss, and the second is the Kullback–Leibler (KL) divergence \cite{kullback1951information}, which acts as a regularizer. \( \beta \) balances these terms \cite{higgins2016beta}. The prior \( p_{\alpha}(z) \) is typically a uniform Dirichlet distribution with parameters \( \alpha \ll 1 \) \cite{burkhardt2019decoupling}. The approximate posterior, \( q(z|x) \), is modeled by a Dirichlet distribution with parameters \( \alpha_p \), derived from the encoder output. The latent vector \( z_i \in \mathbb{R}^{K} \) denotes the document-topic distribution, where \( K \) is the number of topics. \( \phi \in \mathbb{R}^{K \times V} \) represents the normalized topic-word distributions. This approach is fully unsupervised and does not incorporate labels or authors by default.

\subsection{FANToM: Aligning Neural Topic Models}
\label{sec:method}
We now introduce our proposed method, as illustrated in Figure \ref{fig:architecture}. We first explain the integration of labels and authorship information, subsequently merging these elements into a unified framework.

\paragraph{FANToM(L): Aligning Topics with Labels.}
To align topics with labels, we supervise the document-topic distribution using experts\footnote{An expert can be a human labeler or an external source like a large language model (LLM). In our experiments, we use both types of experts.}, ensuring that each document's topics are restricted to its assigned labels. We achieve this by deriving an expert-aligned Dirichlet prior \( p_{\gamma}(z) \) with parameters \( \gamma \), ensuring that the posterior \( q_{\theta}(z|x) \) aligns with this prior through an expert alignment loss.

Let \( \Lambda \) represent the set of possible labels and \( K \) the total number of topics. We define a global topic-label vector \( L \in \{\Lambda \cup \{\text{no-label}\}\}^{K} \), which assigns a label to each topic \( k \). The "no-label" token is used when no label is assigned (e.g., in semi-supervised settings). \( L \) is derived from a mapping \( f: \{k\}_{k=1}^K \rightarrow \{\Lambda \cup \{\text{no-label}\}\} \). For each document \( d \), let \( \lambda^\text{d} \) be the set of labels assigned to the document. We then derive a multi-hot vector \( \mathbb{I}^\text{d} \) as: \( (\mathbb{I}^\text{d}_k = 1 \text{ if } L_k \in \lambda^\text{d}, \text{ else } 0)_{k=1}^K \). The \( k^\text{th} \) element of \( \mathbb{I}^\text{d} \) is 1 if \( L_k \) is a label for the document. We derive the parameters \( \gamma \) for the expert-aligned Dirichlet prior \( p_{\gamma}(z) \) as \( \gamma = \alpha \cdot \mathbb{I}^\text{d} \), where \( \alpha = (\alpha_1, \dots, \alpha_K) \) represents the base Dirichlet distribution over topics.

For example, if \( L = (1, 1, 2, 2, 3) \) and a document \( d \) has labels \( \lambda^\text{d} = \{2\} \), then \( \mathbb{I}^\text{d} = (0, 0, 1, 1, 0) \), and \( \gamma = (0, 0, \alpha_3, \alpha_4, 0) \), ensuring that the document’s topics are restricted to its assigned labels. We illustrate this example extensively in Appendix~\ref{sec:illustrative-ex}. This approach ensures that the model focuses on relevant topics, improving interpretability and topic-label alignment.

\paragraph{FANToM(A): Parameterizing the Topic-Author Distribution.}
The existing VAE-based frameworks do not directly incorporate authorship information. We address this by using a separate decoder to learn a multinomial topic-author distribution based on authorship. Let \( A \) represent the author vocabulary, where each document \( d \) is associated with one or more authors, represented by a multi-hot vector \( a \in \{0,1\}^{|A|} \). The author-decoder, parameterized by \( \psi \), reconstructs the authors from the latent topics. The learned parameter \( \psi \in \mathbb{R}^{K \times |A|} \) represents the topic-author distributions.

\paragraph{Training Objective:}  
Given the expert-aligned prior \(p_{\gamma}(z)\), the observed authors \(\{a_i\}_N\) where \(a_i \in \{0,1\}^{|A|}\), and the author likelihood \(p_{\psi}\), the FANToM training objective is defined as:
\begin{equation}
\label{FANToM}
\begin{split}
    \mathcal{L}_{\text{F}}&\left(\theta, \phi, \psi ; x\right) =-\mathbb{E}_{q_{\theta}\left(z \mid x\right)}\left[\log p_{\phi}\left(x \mid z\right)\right] \\ 
    & \quad \quad \quad \quad \quad -\mathbb{E}_{q_{\theta}\left(z \mid x\right)}\left[\log p_{\psi}\left(a \mid z\right)\right] \\ 
    & \quad \quad \quad \quad \quad +\beta D_{\text{KL}}\left[q_{\theta}\left(z \mid x\right) \| p_{\gamma}\left(z\right)\right],
\end{split}
\end{equation}

where the first term represents the document reconstruction error, the second term accounts for the author reconstruction, and the third term is the expert-alignment loss. The document reconstruction identifies latent topics. The author reconstruction term ensures that the authors associated with a document are constrained by the topics, promoting alignment between authors and topics. The expert-alignment KL loss incorporates the expert-aligned prior \(p_{\gamma}(z)\) instead of \(p_{\alpha}(z)\), enforcing alignment between latent topics and labels. FANToM follows a training regime similar to existing VAE-based NTMs. More details on the training process and the algorithm are provided in Appendix~\ref{sec:training}.

\section{Experiments}
\label{sec:results}
In this section, we first describe the datasets, comparison models, and evaluation metrics used in our experiments. We then demonstrate FANToM's alignment capabilities in Section~\ref{exp:labels} and Section~\ref{sec:authors_align}, followed by a benchmark comparison against baselines in Section~\ref{exp:benchmarking}.

\subsection{Datasets and Preprocessing}
We use four well-known datasets in our experiments. The 20 Newsgroups (20NG) dataset contains around 18,000 newsgroup posts categorized into 20 labels \cite{LANG1995331}. The AG News (AGN) corpus has over a million news articles from 2,000+ sources, divided into four groups \cite{zhang2015character}. The DBpedia-14 (DB-14) dataset consists of 14 distinct classes selected from DBpedia 2014 \cite{zhang2015character}. Lastly, we use the arXiv dataset (arxiv) \cite{arxiv}, which includes six labels and authors with at least ten papers for evaluating author models.

For tokenization, we utilize SpaCy \cite{spacy2}. We remove stop words, punctuation, and words that appear in fewer than 30 documents or in more than 85\% of the documents. Additional details about the datasets and preprocessing are provided in Appendix~\ref{datasets-preprocessing}.

\subsection{Models}

\begin{figure}[t]
    \centering
    \includegraphics[scale=0.62]{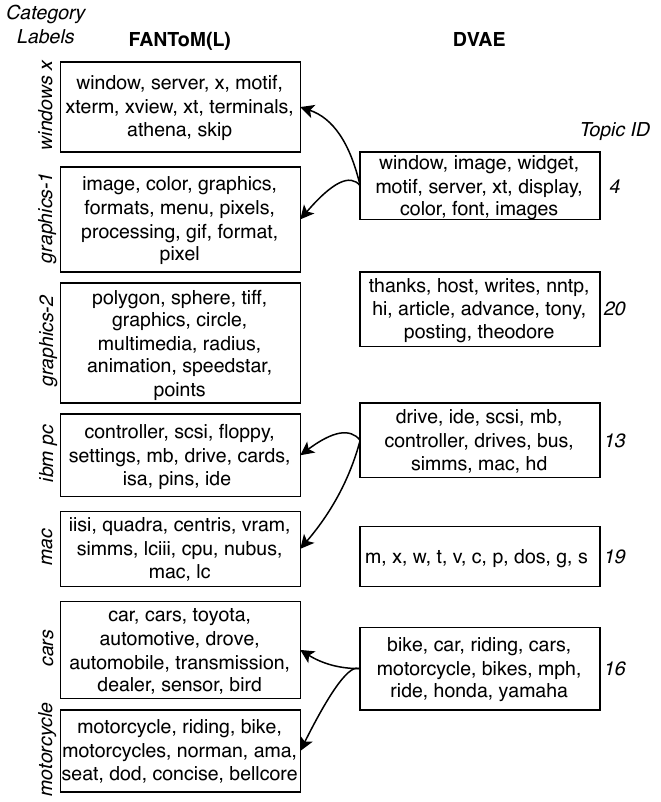}
    \caption{Comparison of topic alignment between FANToM(L) and DVAE (baseline) on the 20NG dataset. The semantically closest topics are linked (right to left). FANToM(L) cleanly separates topics based on labels, while DVAE lacks this distinction. FANToM(L) generates esoteric topics closely aligned with labels and learns multiple topics within the \textit{graphics} label.}
    \label{fig:alignment-topics}
\end{figure} 
\paragraph{Baselines:} 
We compare FANToM against several supervised and unsupervised topic models. As supervised baselines, we use SCHOLAR \cite{card-etal-2018-neural}, TAM \cite{wang2020neural}, and Labeled LDA (L-LDA) \cite{ramage-etal-2009-labeled}. For author modeling, we include the Author-Topic Model (ATM) \cite{zvi2004}.\footnote{We use implementations from \cite{bab2min} and \cite{rehurek_lrec} for L-LDA and ATM, respectively.} For unsupervised NTMs, we include Dirichlet-VAE (DVAE) \cite{burkhardt2019decoupling}, Embedded Topic Model (ETM) \cite{dieng2020topic}, and ECRTM \cite{wu2023effective}, which extends ETM. Additionally, we use Contextualized Topic Models (CTM) \cite{bianchi2020cross}, with document embeddings from SBERT \cite{reimers-gurevych-2019-sentence}.

\paragraph{FANToM:} Our proposed framework can be integrated into existing NTMs. We present FANToM variants for all NTM baselines, including SCHOLAR, DVAE, ETM, ECRTM, and CTM. To maintain consistency, all neural baselines and FANToM use a Dirichlet prior. See Appendix~\ref{modeling-details} for more details.

\setlength{\tabcolsep}{0.90pt}
\begin{table*}[ht]
    \centering
    \begin{tabular}{c|c|l|l|l|l|l|l|l|l|l|l|l|l}
    \hline \hline 
        &Models & \multicolumn{3}{c|}{20NG} & \multicolumn{3}{c|}{AGN} & \multicolumn{3}{c|}{DB-14} & \multicolumn{3}{c}{arxiv} \\
        & & TQ & Purity & NMI & TQ & Purity & NMI & TQ & Purity & NMI & TQ & Purity & NMI\\
        
         \hline
         \multirow{7}{*}{\rotatebox{90}{baseline}} & L-LDA & $0.192$ & $0.629$ & $\textbf{0.681}$ & $0.320$ & $0.815$ & $0.747$ & $0.518$ & $0.894$ & $0.876$ & $0.246$ & $0.875$ & $0.806$\\
         & SCHOLAR & $0.391$ & $0.341$ & $0.375$ & $0.372$ & $0.617$ & $0.581$ & $0.601$ & $0.811$ & $0.809$ & $0.343$ & $0.796$ & $0.779$\\
         & TAM & $0.377$ & $0.328$ & $0.361$ & $0.385$ & $0.603$ & $0.570$ & $0.620$ & $0.785$ & $0.832$ & $0.355$ & $0.770$ & $0.791$\\
         & DVAE & $0.354$ & $0.336$ & $0.325$ & $0.393$ & $0.743$ & $0.691$ & $0.628 ^\ddag$ & $0.829$ & $0.810$ & $0.339$ & $0.862$ & $0.836$\\
         & ETM & $0.398$ & $0.491$ & $0.482$ & $0.310$ & $0.795$ & $0.774$ & $0.603$ & $0.902$ & $0.896$ & $0.264$ & $0.884$ & $0.875$\\
         & CTM & $0.412$ & $0.390$ & $0.374$ & $0.371$ & $0.763$ & $0.749$ & $0.603$ & $0.858$ & $0.841$ & $0.324$ & $0.814$ & $0.799$\\
         & ECRTM & $0.434$ & $0.628$ & $0.617$ & $0.391$ & $0.826$ & $0.809$ & $0.640$ & $0.894$ & $0.881$ & $0.315$ & $0.886$ & $0.862$\\
         \hline
         \multirow{4}{*}{\rotatebox{90}{FANToM(ours)}} & SCHOLAR & $0.403 ^\ddag$ & $0.649 ^\ddag$ & $0.601 ^\ddag$ & $0.397 ^\ddag$ & $0.841 ^\ddag$ & $0.819 ^\ddag$ & $0.608$ & $0.915 ^\ddag$ & $0.887 ^\ddag$ & $\textbf{0.375} ^\ddag$ & $0.919 ^\ddag$ & $0.893 ^\ddag$\\
         & DVAE & $0.424 ^\ddag$ & $0.629 ^\ddag$ & $0.611 ^\ddag$ & $\textbf{0.409} ^\ddag$ & $0.837 ^\ddag$ & $0.821 ^\ddag$ & $0.611$ & $0.898 ^\ddag$ & $0.874 ^\ddag$ & $0.336$ & $0.910 ^\ddag$ & $0.904 ^\ddag$\\
         & ETM & $0.399$ & $0.670 ^\ddag$ & $0.641 ^\ddag$ & $0.391 ^\ddag$ & $0.854 ^\ddag$ & $0.830 ^\ddag$ & $0.630 ^\ddag$ & $0.924$ & $0.905$ & $0.310 ^\ddag$ & $0.920 ^\ddag$ & $0.915 ^\ddag$\\
         & CTM & $0.436 ^\ddag$ & $0.651 ^\ddag$ & $0.617 ^\ddag$ & $0.395 ^\ddag$ & $0.847 ^\ddag$ & $0.828 ^\ddag$ & $0.622 ^\ddag$ & $0.916 ^\ddag$ & $0.908 ^\ddag$ & $0.342 ^\ddag$ & $0.895 ^\ddag$ & $0.873 ^\ddag$\\
         & ECRTM & $\textbf{0.441}$ & $\textbf{0.710} ^\ddag$ & $0.681 ^\ddag$ & $0.402$ & $\textbf{0.918} ^\ddag$ & $\textbf{0.873} ^\ddag$ & $\textbf{0.651} ^\ddag$ & $\textbf{0.937} ^\ddag$ & $\textbf{0.917} ^\ddag$ & $0.332 ^\ddag$ & $\textbf{0.941} ^\ddag$ & $\textbf{0.925} ^\ddag$\\
    \hline
    \end{tabular}
    \caption{Comparison of the proposed FANToM(L) (bottom) against the respective non-aligned baselines and L-LDA (top) using Topic Quality, Purity, and NMI measures. The best results across datasets are highlighted in bold, and significantly better results (p-value < 0.05) between baselines and FANToM(L) are marked with $^\ddag$. Our FANToM approach significantly outperforms the existing baselines.}
    \label{tab:benchmarking}
\end{table*}

\subsection{Evaluation Measures}
\label{sec:eval}
In addition to qualitative analysis, we evaluate model performance quantitatively using the standard Topic Quality (TQ) measure, which is defined as the product of Topic Coherence (TC) and Topic Diversity (TD) \cite{dieng2020topic}. TQ reflects both interpretability and diversity of topics. TC, calculated via the $C_V$ coherence score \cite{roder2015exploring}, measures the co-occurrence of top words within a topic using a reference corpus. We use the WikiText-103 dataset \cite{merity2016pointer} as our reference corpus, consisting of approximately 2 million Wikipedia articles. TD assesses the proportion of unique words across all topics, with scores close to zero indicating redundancy and scores near one reflecting high diversity. Additionally, we perform experiments on document clustering to assess the topic alignment using Purity and NMI, following \cite{wu2023effective,hoyle2022neural}. Purity assesses topic homogeneity by assigning each topic the most frequently co-occurring label, while NMI evaluates the mutual information between true labels and predicted topics, thereby measuring alignment in terms of precision and recall. Moreover, across all experiments, we conduct a two-tailed t-test to assess significance.

\subsection{Alignment of Topics with Labels}
\label{exp:labels}
To demonstrate the problem of misaligned topics, we compare the alignment of FANToM(L) with the DVAE baseline on the 20NG dataset (Figure~\ref{fig:alignment-topics}). DVAE tends to generate general, noisy topics, prioritizing common words and leading to information loss. For example, in Topic ID 4, DVAE merges the graphics and windows topics, adding noisy words like "font" that do not align with the label. In contrast, FANToM(L) separates topics more effectively, guided by labels. For instance, DVAE fails to generate a distinct \textit{mac} topic, while FANToM(L) successfully separates it. FANToM(L) produces more expressive, homogeneous topics closely aligned with the labels.

To capture multiple topics within a single label, we assigned two topic indices to the \textit{graphics} label using the topic-label vector $L$, resulting in two distinct topics (\textit{graphics-1} and \textit{graphics-2}). As seen in Figure \ref{fig:alignment-topics}, FANToM(L) distinguishes between topics related to digital graphics software (\textit{graphics-1}) and geometry (\textit{graphics-2}). These distinctions are weak or absent in DVAE.

To quantify alignment, we calculate purity and NMI \cite{wu2023effective} to measure how well topics in the latent space correspond to assigned labels. As shown in Table~\ref{tab:benchmarking}, FANToM(L) significantly outperforms baselines in producing more aligned document-topic distributions, as indicated by the higher purity and NMI scores.

\subsection{Alignment of Topics with Authors}
\label{sec:authors_align}

Authorship information can further refine topic alignment. As shown in Table~\ref{tab:author-align}, FANToM(A) outperforms ATM, the baseline author-topic model. Additionally, FANToM, which aligns both labels and authors, achieves higher scores than FANToM(L) and FANToM(A) individually. For this experiment, we use the arXiv dataset, the only one in our benchmarks with author information. Figure~\ref{fig:main-fig} illustrates topic-author alignment, with additional examples provided in Appendix \ref{sec:authors_align_ap}.

\subsection{Benchmarking FANToM(L)}
\label{exp:benchmarking}
To measure the quality of generated topics, we benchmark the proposed label alignment models against state-of-the-art NTMs in Table~\ref{tab:benchmarking}. Since FANToM can be integrated with any existing VAE-based topic model variant, we incorporate FANToM with all the baselines in the benchmarking. Labels from the datasets are used to train our models, and to ensure consistency, the total number of topics learned across all models is set to match the number of labels. Our results are averaged over five independent runs. Contrary to our expectation of a trade-off between alignment and topic quality, incorporating labels does not harm the quality of the topics produced; on the contrary, it even improves the quality significantly. Individual coherence and diversity measures are listed in Appendix \ref{sec:quantitative_eval} for further analysis.

We also conduct benchmarking experiments in a semi-supervised scenario where we do not have labels for all the topics in our corpus. To address this, we assign the "no-label" designation to all documents lacking labels. In our experiment, we randomly remove labels from 50\% of the documents across all datasets and learn 50 and 200 topics, resulting in some topics being labeled while the rest are labeled as "no-label." FANToM consistently outperforms the baseline models, as shown in Table~\ref{tab:semi-sup} (Appendix~\ref{sec:quantitative_eval}).

\setlength{\tabcolsep}{10pt}
\begin{table}[t!]
    \centering
    \begin{tabular}{c|c|c|c}
    \hline \hline
    Models & TQ & Purity & NMI  \\
    \hline
        ATM &  $0.287$ & $0.730$ & $0.759$\\
        DVAE &  $0.339$ & $0.862$ & $0.836$\\
        FANToM(L) &  $0.336$ & $0.910$ & $0.904$\\
        FANToM(A) &  $0.354$ & $0.889$ & $0.861$\\
        FANToM & $\textbf{0.362}$ & $\textbf{0.951}$ & $\textbf{0.948}$\\
    \hline
    \end{tabular}
    \caption{Comparison of FANToM (L), FANToM (A), and FANToM against non-aligned baseline (DVAE) and aligned statistical baseline (ATM). Best results are highlighted in bold. Both FANToM and FANToM (A) significantly outperform ATM on all aspects.}
    \label{tab:author-align}
\end{table}

\begin{figure*}[t]
    \centering
    \includegraphics[scale=0.9]{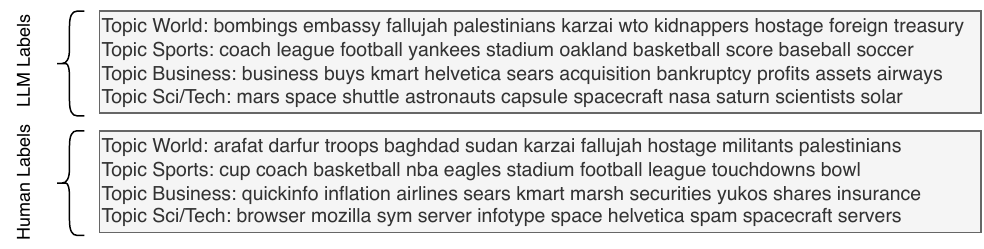}
    \caption{Comparison between the topic words estimated using human labels (bottom) and LLM labels (top) in the ag news corpus, using FANToM(L) for topic estimation. Both align well with the corresponding labels.}
    \label{fig:human_llm}
\end{figure*}

\begin{figure}[t]
    \centering
    \includegraphics[scale=0.55]{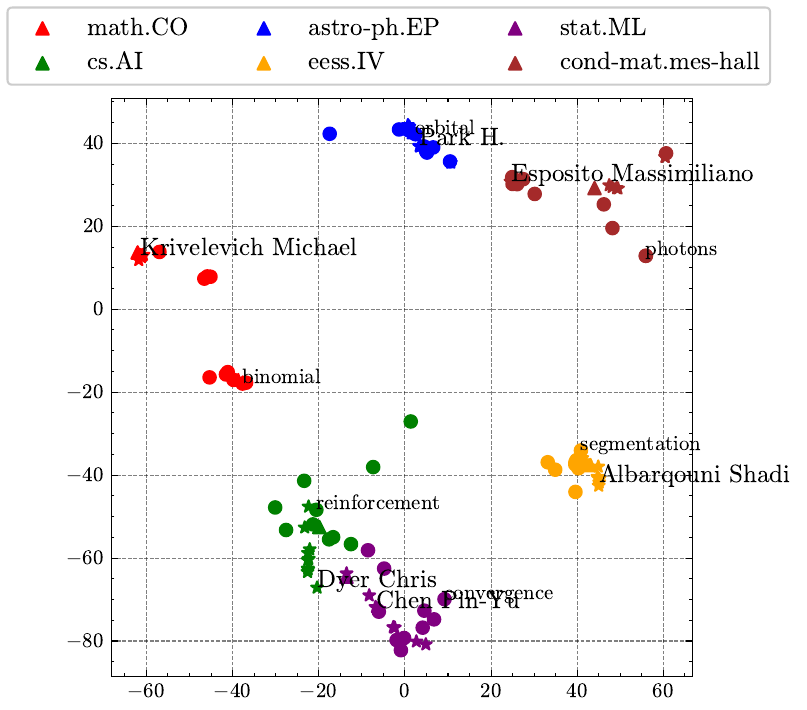}
    \caption{The shared embedding space between author ($\star$), word ($^\ddag$), and topic ($\blacktriangle$) embeddings shows authors and words in close proximity to their respective topics.}
    \label{fig:embedding}
\end{figure}

\section{Discussion}
\label{sec:discuss}

Our experiments confirm the achieved alignment qualitatively and quantitatively. We do not observe any trade-off between alignment and topic quality; on the contrary, we observe improved quality compared to the SOTA baselines. Crucially, we also show that the statistical topic model Labeled LDA performs worse than all neural topic models, even though Labeled LDA is trained with the Gibbs sampling algorithm, which is less prone to local minima during training than variational inference. This demonstrates that aligning neural topic models is a crucial step towards further improving their performance and practical relevance. We now discuss some applications of FANToM's alignment.

\paragraph{Learning an Embedding Space Between Topics and Authors:} We are able to extract informative word, topic, and author embeddings by using FANToM with embedding (ETM) decoders for both topics and authors. Figure~\ref{fig:embedding} shows the TSNE \cite{van2008visualizing} projection of the embeddings in a shared embedding space. We assign the learned label to each topic, allowing us to easily identify topics. The figure demonstrates that words and authors belonging to a particular topic are close to each other, while each topic cluster remains distinct from others. The learned embeddings further open avenues for exploration and analysis of associated authors and topics.

\paragraph{Using LLMs as Experts}

Labeling datasets can be challenging, but Large Language Models (LLMs) offer an alternative for assigning labels to unlabeled documents. We explore this approach using BART \cite{lewis2019bart} for zero-shot classification, leveraging a pretrained model \cite{wolf2019huggingface} fine-tuned on Natural Language Inference (NLI) \cite{N18-1101} following \cite{yin2019benchmarking}. This method frames classification as an NLI task, where a sequence serves as the premise and each potential label is tested via a hypothesis (e.g., “This text is about sports”). By computing entailment and contradiction probabilities, we derive label assignments. Applying this to our dataset and comparing with human labels, we achieve 92\% accuracy, validating LLM effectiveness in document categorization. Additionally, using LLM-generated labels for the AG News corpus (AGN) with FANToM(L) yields a mean topic quality of 0.503, outperforming the 0.409 obtained with human labels. Figure \ref{fig:human_llm} illustrates the topic words estimated from both sources, showing strong alignment with their corresponding labels.

\paragraph{Author Interests and Similarity:}
In addition to aligned topics, we also learn authors' interests as FANToM maps each author to a topic distribution and vice versa. We first validate this comparing FANToM's recommended labels for each author to ground truth labels. We find that FANToM accurately associates 91.6\% of authors with their respective labels.

\begin{figure}
    \centering
    \includegraphics[scale=0.8]{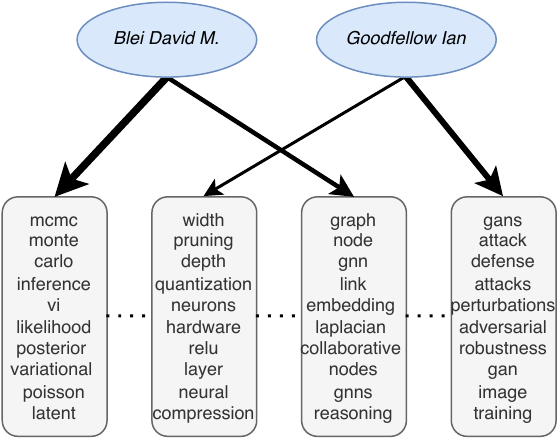}
    \caption{The top two topics of \textit{Blei David M.} and \textit{Goodfellow Ian}, based on cosine similarity, accurately represent the respective research interests of the authors, aligning well with their contributions in probabilistic modeling and generative adversarial networks, respectively.
    }
    \label{fig:author_topics}
\end{figure}

To evaluate author interests, we train the proposed FANToM(A) model on the "stats.ML" label from the \cite{arxiv} corpus, spanning the years 2010 to 2020. Each author is associated with different topics (interests), as illustrated in Figure~\ref{fig:author_topics}, where the line thickness corresponds to the weight of the association. We specifically select two prominent researchers in the field, namely \textit{Blei David M.} and \textit{Goodfellow Ian}, and identify their top two matching topics. The figure illustrates the correct alignment between these researchers and their respective research interests. Furthermore, the author-topic vectors can also be used to compute the similarity between authors. One such similarity matrix is depicted in Figure \ref{fig:similarity} in Appendix~\ref{add-eval}.

\paragraph{Topic Models and LLMs:}  
In recent years, the combination of topic modeling with LLMs has emerged as a key area of research. Recent efforts have explored combining topic modeling with LLMs, such as TopicGPT \cite{pham-etal-2024-topicgpt}, which uses prompt engineering to address topic modeling and does not provide a framework for alignment. However, the direct prompting method incurs high computational costs for processing prompts of an individual document. FANToM, on the other hand, employs a VAE for topic modeling and uses a smaller, task-specific LLM for labeling, offering alignment, reduced computational costs, greater flexibility and efficiency. We discuss this aspect in detail in the Appendix \ref{sec:llms-tm}.

\section{Conclusion}
In this work, we propose FANToM, a novel neural architecture for aligning Neural Topic Models (NTMs) with expert-assigned labels and authorship information. Our results show that (i) the alignment effectively captures the corresponding associations among topics, labels, and authors, and (ii) FANToM outperforms existing state-of-the-art models in terms of topic quality. This underscores the importance of aligning topic models and paves the way for potential downstream applications. We investigate learning a shared embedding space between topics, authors, and words, enabling us to identify author interests and compute the similarity between and within authors and topics. As incorporating prior knowledge into machine learning has recently seen significant success \cite{nagda2024setpinns,nagdapits,specht2024hannahardconstraintneuralnetwork,vollmer-2024,jirasek2023workshop,manduchi2024challenges}, promising lines of future research include the possibility of combining prior knowledge with the modeling of joint topic, label, and author hierarchies.

\section*{Limitations}
Our approach may face challenges when an author suddenly writes about a completely unrelated topic. However, such cases are relatively rare, and when there is even partial thematic overlap, we believe FANToM remains robust. Addressing extreme cases of topic divergence presents an interesting direction for future research. Although FANToM is a flexible framework, it necessitates the use of labels and authors for alignment, which involves collecting metadata. When expert labels are not readily available, using even a smaller LLM as an alternative can result in higher computational costs and resource demands. Additionally, LLMs face limitations due to context length constraints. In our experiments, truncation was required, but it is not an optimal solution. Future research could explore methods to handle full documents within these length limits, such as processing documents in chunks, selecting representative segments, or summarizing the content. Another challenge with LLMs is multilinguality; for some languages, finding a small-scale LLM for label assignment may be impractical and increase reliance on human experts for label assignment.

\section*{Ethical Statement}

In this study, we incorporated authorship information into topic models while ensuring ethical considerations. To acquire the dataset, we obtained it from the openly available \cite{arxiv}, which is licensed under the \textit{CC0 1.0 Universal (CC0 1.0) Public Domain Dedication}, allowing us unrestricted use of the data. We specifically utilized the abstracts and author names of openly available papers. It is important to note that we have mentioned author names as they were presented to us by our model, without making any alterations. We did not have access to any sensitive or restricted information.

\section*{Acknowledgments}
The authors acknowledge support by the Carl-Zeiss Foundation, DFG awards BU 4042/2-1, FE 2282/6-1, and FE 2282/1-2, as well as the BMBF award 01|S2407A.

\bibliography{custom}

\appendix

\section{Modeling Details}
\label{modeling-details}

Let $x$ be the BoW representation of a document. Let $V$ and $K$ be vocabulary size and no of topics respectively. The common encoder used across different models is:

Encoder:
\begin{equation*}
\begin{split}
X_{1} &= \text{ReLU}(L^{1}\cdot x + L_{\text{bias}}^{1}) \\
X_{2} &= \text{dropout}(X_{1}, p_{\text{keep}}) \\
X_{3} &= \text{BatchNorm}(L^{2}\cdot X_{2} + L_{\text{bias}}^{2}) \\
\text{enc\_out} &= \text{Softplus}(X_{3}),
\end{split}
\end{equation*}

A common linear decoder with Dirichlet prior is given as:

Decoder:
\begin{equation*}
\begin{split}
    X_{4} &= \text{BatchNorm}(L^{3}\cdot z + L_{\text{bias}}^{3})\\
    \text{recon} &= \text{LogSoftmax}(X_{4})
\end{split}
\end{equation*}

Certain models (such as ETM) use an embedding decoder. $\alpha$ and $\delta$ represent topic and word embeddings, respectively:

Decoder ETM:
\begin{equation*}
\begin{split}
    X_{4} &= z \cdot \eta \cdot \delta\\
\text{recon} &= \text{LogSoftmax}(\text{BatchNorm}(X_{4}))
\end{split}
\end{equation*}

For FANToM we have two decoders for text and authors of the document.

Decoder FANToM:

\begin{equation*}
\begin{split}
X_{4} &= \text{BatchNorm}(L^{3}\cdot z + L_{\text{bias}}^{3})\\
 X_{5} &= \text{BatchNorm}(L^{4}\cdot z + L_{bias}^{4})\\
\text{recon\_doc} &= \text{LogSoftmax}(X_{4}) \\
\text{recon\_author} &= \text{LogSoftmax}(X_{5})
\end{split}
\end{equation*}

The hyperparameters used in the experiments are given in Table \ref{tab:hyperparameters}.

Every model, including its FANToM variant, is constructed as follows:

\paragraph{SCHOLAR:}
SCHOLAR utilizes the common Encoder with the linear Decoder. The softmax of the output of the Encoder, denoted as softmax(enc\_out), is used as input to a separate classifier, which predicts document labels.

\paragraph{DVAE:}
DVAE employs the common Encoder with the linear Decoder.

\paragraph{ETM:}
ETM utilizes the common Encoder with an embedding decoder (Decoder ETM). The word embeddings are initialized with GloVe embeddings.

\paragraph{CTM:}
CTM employs the common Encoder with the linear Decoder. The distinction lies in the input to the encoder, which consists of contextualized embeddings rather than Bag of Words (BoW). The contextualized embeddings are sourced from the SBERT model.

\paragraph{ECRTM:}
ECRTM employs the common Encoder with the ETM Decoder. The distinction lies in the optimization function, where we use implementation provided by the authors\footnote{https://github.com/BobXWu/ECRTM}.

All of the aforementioned models are also constructed with their corresponding FANToM variants. In these variants, only the priors are manipulated in the objective function (as mentioned in described in Section \ref{sec:method} of the main paper), and a separate author decoder is introduced to learn the topic-author multinomial distribution. The core architecture remains unchanged in all cases.

This also confirms the adaptability of our architecture to seamlessly incorporate any VAE-based topic model. Furthermore, it demonstrates the ability to align the learned topics with expert-assigned labels, and to leverage author information for establishing a meaningful correspondence between topics, authors, and labels.

\setlength{\tabcolsep}{13pt}
\begin{table*}[!ht]
\caption{Hyperparameter settings for the experiments.}
\label{tab:hyperparameters}
\begin{center}
\begin{tabular}{l|c}
\hline
\hline
Hyperparameter & Value \\
\hline
batch size & 128\\
$\alpha$ & 0.02\\
$\beta$ & 2 \\
Learning Rate & 0.001\\
Max Epochs & 100\\
$p_{keep}$ & 0.25\\
$L^1$ & $\mathbb{R}^{\text{vocab\_size $\times$ 512}}$\\
$L^2$ & $\mathbb{R}^{\text{512 $\times$ total\_topics}}$\\
$L^3$ & $\mathbb{R}^{\text{total\_topics $\times$ vocab\_size}}$\\
$L^4$ & $\mathbb{R}^{\text{total\_topics $\times$ author\_size}}$\\
$\delta$ & $\mathbb{R}^{\text{total\_words $\times$ 300}}$\\
$\eta$ & $\mathbb{R}^{\text{total\_words $\times$ 300}}$\\
word embeddings & \cite[GloVe]{pennington2014glove}\\
train:val:test & 70:15:15 \\
\hline
\end{tabular}
\end{center}
\end{table*}

\section{FANToM}
\begin{figure}
    \centering
    \includegraphics[width=0.5\linewidth]{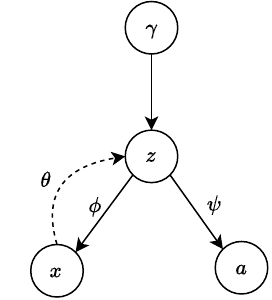}
    \caption{The graphical model of our framework is also shown, where solid lines represent the generative distribution $p$, and dotted lines represent the variational distribution $q$.}
    \label{fig:graphical-model}
\end{figure}
\subsection{Illustrative example to construct topic-label vector}
\label{sec:illustrative-ex}
Consider a document \( d \) in a corpus with five topics (\( K = 5 \)). We define a topic-label linking vector \( L_k = (\text{sport}, \text{cars}, \text{weather}, \text{no-label}, \text{no-label}) \), where each element links a topic index to a corresponding label. If the label for document \( d \) is "sport", we define a label indicator vector \( \mathbb{I}^\text{d} = (1, 0, 0, 0, 0) \), indicating that only the first topic is active. If no labels are provided, the indices corresponding to "no-label" are activated, resulting in \( \mathbb{I}^\text{d} = (0, 0, 0, 1, 1) \). For cases where multiple labels are assigned, such as both "sport" and "cars", the indicator vector becomes \( \mathbb{I}^\text{d} = (1, 1, 0, 0, 0) \).

Using this label indicator vector, the modified prior parameter is defined as \( \gamma = \alpha \cdot \mathbb{I}^\text{d} \), where \( \alpha \) is a scaling factor. This allows for flexible topic assignments based on the available labels. For example, if \( L_k =\) (sport, sport, cars, cars, weather, weather, no-label, no-label) , multiple topics may be linked to a single label. This allows the model to learn multiple focused topics.

\subsection{Training FANToM}
\label{sec:training}
The training algorithm for FANToM, outlined in Algorithm \ref{alg:a-ntm}, begins by taking as input a document set $D$, an expert $E$, a prior parameter $\alpha$, and topic-label assignments $L_k$. The process starts with the expert providing labels \( l \) for the documents, which are then used to create a multi-hot vector \( \mathbb{I}^\text{d} \) that reflects the association between labels and topics.

The algorithm processes batches \( \mathcal{B} \) of documents from the set \( D \). For each batch, the bag-of-words (BoW) representation \( x \) and an author representation \( a \) are extracted. An encoder network, parameterized by \( \theta \), computes the posterior parameters \( \alpha_{p} \) from the BoW representation. Simultaneously, a prior distribution is constructed using the expert-provided labels, with parameters given by \( \gamma = \alpha \cdot \mathbb{I}^\text{d} \).

A variable \( z \) is then sampled from a Dirichlet distribution \( \text{Dirichlet}(\alpha_{p}) \), representing the document-topic distribution. This sampled \( z \) is input into two decoder networks: one reconstructs the document as \( x' \), while the other reconstructs the author representation as \( a' \).

The model parameters are updated by minimizing the objective function defined in Eq.~\ref{FANToM}. This iterative process of batch processing, parameter updating, and reconstruction continues until the algorithm converges, ensuring the model effectively learns both document and author representations aligned with the provided topic labels.

Throughout the training, the interaction between the posterior parameters derived from the encoder, the prior informed by expert labels, and the outputs from the decoders continually refines the model’s understanding of topic, label, and author associations. This optimization process ensures that FANToM captures the underlying topic structures in the document set \( D \), aligning them with expert-provided labels and authors.

\begin{algorithm}[tb]
  \caption{Training FANToM}
   \label{alg:a-ntm}
\begin{algorithmic}[1]
   \STATE {\bfseries Input:} Documents $D$, Expert $E$, Prior parameter $\alpha$, Topic-label assignments $L_k$
   \STATE $\lambda\leftarrow$ Get labels from expert $E$ for documents $D$
   \STATE $\mathbb{I}^\text{d} \leftarrow$ Create multi-hot vector based on $\lambda^\text{d}$ and $L_k$ for $d \in D$
   \STATE Initialize model parameters $(\theta, \phi, \psi)$
   \WHILE{not converged}
   \FOR{batch $\mathcal{B}$ in $D$}
   \STATE Extract BoW $x$, authors $a$, and multi-hot vector $\mathbb{I}^\text{d}$ from $\mathcal{B}$
   \STATE $\gamma \leftarrow \alpha \cdot \mathbb{I}^\text{d}$ (modify prior parameter)
   \STATE $\alpha_{p} \leftarrow \text{Encoder}(\theta;x)$ (encode words)
   \STATE Sample $z \sim \text{Dirichlet}(\alpha_{p})$
   \STATE $x' \leftarrow \text{Decoder}(\phi;z)$ (reconstruct document)
   \STATE $a' \leftarrow \text{Decoder}(\psi;z)$ (reconstruct author)
   \STATE Compute gradients $\nabla_{(\theta, \phi, \psi)} L_{\text{F}}(\theta,\phi,\psi;x, \gamma)$ (Eq.~\ref{FANToM})
   \STATE Update parameters $(\theta, \phi, \psi)$
   \ENDFOR
   \ENDWHILE
\end{algorithmic}
\end{algorithm}

\section{Datasets and Preprocessing}
\label{datasets-preprocessing}

For dataset preparation, we use SpaCy \cite{spacy2} for data tokenization. Additionally, we eliminate common stop words, punctuations, as well as high and low-frequency words. High and low-frequency words are determined by excluding words that appear in over 85\% of the documents or in fewer than 30 documents. This standardization is applied across all models.

\section{Using LLMs as Expert}
\label{llmexpert}

Not all datasets come with accompanying labels and labeling every corpus can be a challenging task. However, an alternative approach involves using Large Language Models (LLMs) \cite{ostheimer-etal-2023-call, ostheimer2024text} to assign labels to unlabeled documents. In our study, we explore this possibility by using the capabilities of BART \cite{lewis2019bart} for zero-shot text classification. To achieve this, we use a pretrained model from \cite{wolf2019huggingface}, which undergoes fine-tuning on Natural Language Inference (NLI) \cite{N18-1101} following the methodology described by \cite{yin2019benchmarking}.

This method operates by treating the sequence to be classified as the NLI premise and constructing a hypothesis for each potential label. For example, when determining if a sequence pertains to the "sport" label, we can formulate a hypothesis such as "This text is about sports." By converting the probabilities for entailment and contradiction, we obtain the label probabilities.

We apply this approach to our selected dataset and supply prospective labels from the label set. By comparing the results obtained through this method with the available human labels, we achieve an accuracy of 92\%. This validation further supports the practicality and effectiveness of using LLMs as experts for document categorization. Furthermore, we use the LLM labels of the ag news corpus (AGN) and use FANToM(L) to estimate topics. The results demonstrate a mean topic quality of 0.503, surpassing the 0.409 achieved through human labeling. Figure \ref{fig:human_llm} provides an illustration of the topic words estimated by both human and LLM labels. Notably, both sets of topics align well with the corresponding labels.


\section{Runtime}
\label{runtime}
In this section, we present the runtime of the proposed models compared to existing models. All experiments were conducted on NVIDIA V100 GPUs. The runtimes are presented in Table \ref{tab:runtime}. The results demonstrate that the proposed methods have minimal impact on the overall runtime. It is important to note that only the model training time is considered in the runtime analysis, while data preparation time is excluded. The incorporation of labels and authors introduces a data preparation threshold, but it remains constant and does not scale with the model's runtime.

\section{Additional Evaluation}
\label{add-eval}
\begin{figure*}
    \centering
    \includegraphics[scale=0.75]{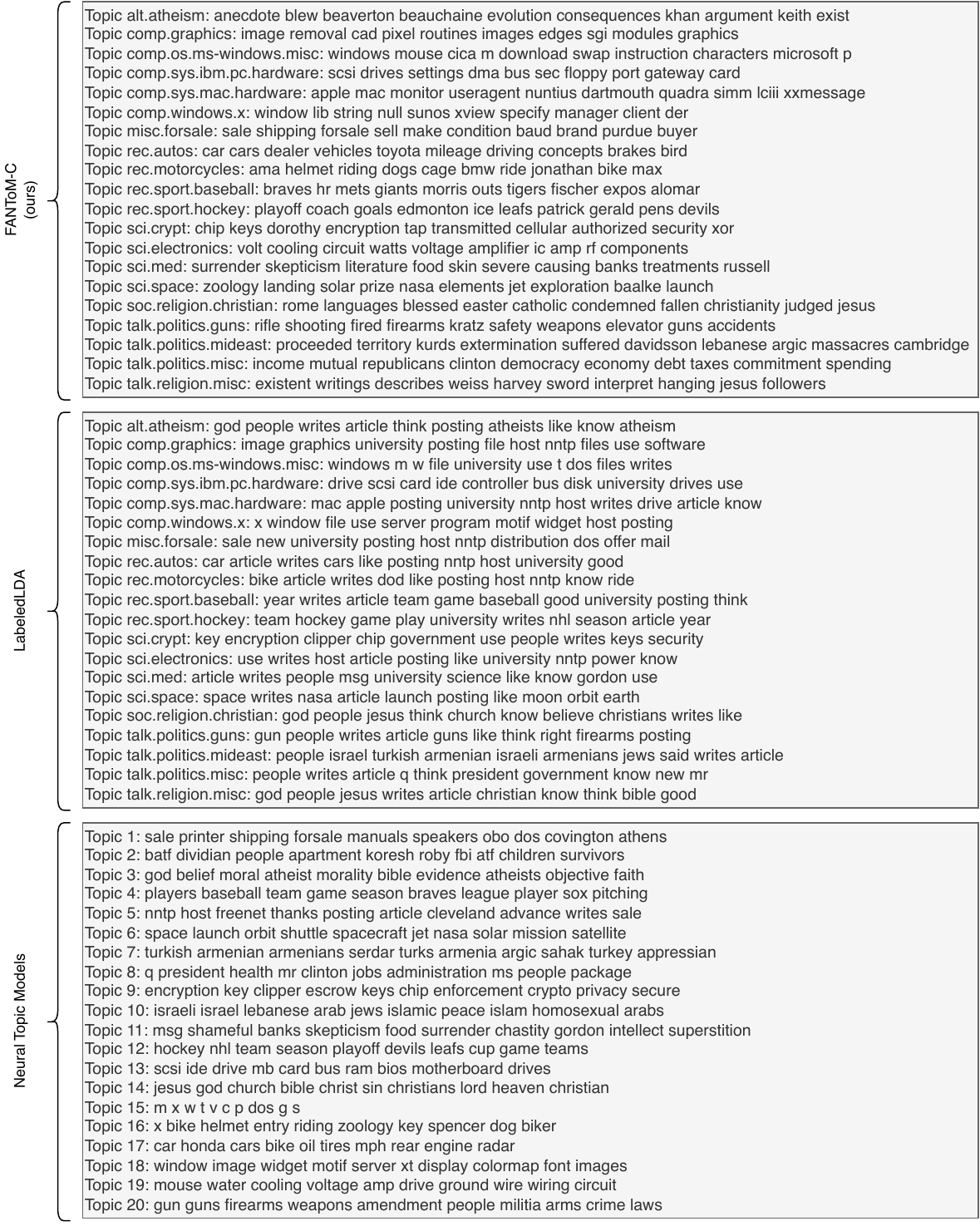}
    \caption{This figure compares FANToM(L) (top), Labeled LDA (middle), and Neural Topic Models (bottom) on the 20NG dataset. FANToM(L) exhibits aligned and esoteric topics consistent with human intentions, while Labeled LDA shows repetitive generic words in many topics (e.g., "host," "writes," "articles"), leading to low diversity. Neural Topic Models do not have association with provided labels and fail to produce certain topics completely, unlike FANToM(L) (e.g., "mac", "graphics", "electronics").}
    \label{fig:labeled_topics}
\end{figure*}

Apart from the main paper's evaluation, we provide supplementary assessments. Figure \ref{fig:labeled_topics} showcases extended qualitative analysis by revealing all the topics acquired by Labeled LDA, existing NTMs, and our proposed FANToM on the 20ng dataset. This illustration highlights the concordance with labels and also showcases the esoteric topics uncovered by our method. It further underscores the suboptimal quality of topics obtained through statistical models.

\subsection{Additional Quantitative Evaluation}
\label{sec:quantitative_eval}

Tables \ref{tab:benchmarking-coherence} and \ref{tab:benchmarking-diversity} delve into topic coherence and diversity in comparison to baselines, thus demonstrating the superior quality of topics learned by our model.

We also validate alignment quantitatively by conducting text classification using FANToM(L). We use the encoder to extract posterior probabilities for the test set during inference. We evaluate each model based on Top-3 and Top-5 accuracy and F1 scores. As demonstrated in  Tables~\ref{tab:classification} and \ref{tab:classification-acc}, FANToM(L) yields good classification and F1 scores which suggests alignment with the expert-assigned labels.

We also conduct benchmarking experiments in a semi-supervised scenario where not all topics in the corpus have labels. To manage this, we designate the "no-label" category for documents lacking labels. In our experiment, we randomly remove labels from 50\% of the documents across all datasets and train models to learn 50 and 200 topics. This setup results in a mixture of labeled and "no-label" topics, reflecting real-world situations where not all documents may have known labels. Analyzing the results, shown in Table~\ref{tab:semi-sup}, FANToM consistently outperforms the baseline models, demonstrating its robustness even in the absence of complete label information. Specifically, the FANToM variants show higher topic coherence (TC) and topic quality (TQ) across datasets. Despite the removal of labels, FANToM maintains superior peformance and consistently achieves better overall scores, solidifying its advantage in semi-supervised settings.

\subsection{Topic-Author Alignment}
\label{sec:authors_align_ap}
\begin{figure*}
    \centering
    \includegraphics[scale=1.0]{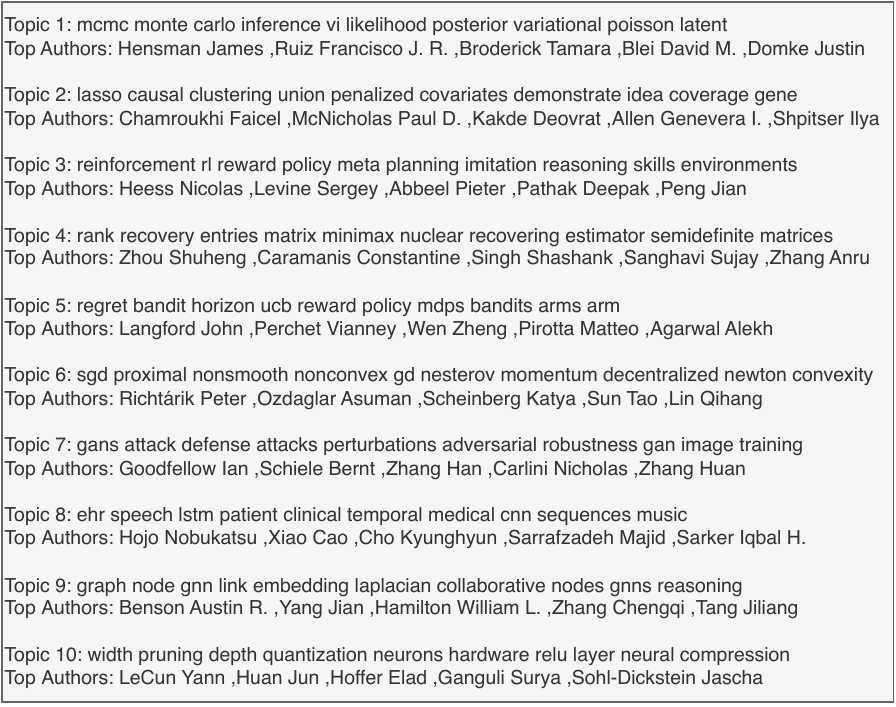}
    \caption{Cosine Similarity Matrix illustrating the relationships between prominent ML authors, determined by analyzing their associated topic vectors through the FANToM(A) model. High similarity scores indicate shared research areas among authors, revealing clusters of researchers with overlapping contributions and thematic focus in machine learning. It can provide insights into academic collaborations, interdisciplinary influences, and the evolution of research trends within the field.}
    \label{fig:author_topics_ml}
\end{figure*}

Our framework allows us to link topics with authors using two distinct approaches: FANToM, which uses labels, and FANToM(A), which does not. In Table~\ref{tab:authors}, we present results that demonstrates that our method associates topics with authors and labels from the arxiv dataset. The authors associated with the topics belong to their respective field of research. Including labels allows us to validate this association by checking the respective topic authors within their labels. Further we also list results from FANToM(A) on stats.ML subset from the arxiv dataset in Figure \ref{fig:author_topics_ml}.


\section{Discussion on Topic Models and LLMs}
\label{sec:llms-tm}

In recent years, combining topic modeling with large language models (LLMs) has emerged as a key area of research, aiming to harness the strengths of both techniques. 

Traditionally, topic modeling has employed methods like Latent Dirichlet Allocation (LDA) and its variants to uncover hidden topics in a corpus based on word co-occurrence patterns. However, these methods often face challenges with interpretability and coherence.

The introduction of Variational Autoencoders (VAEs) brought a new level of flexibility to topic modeling by incorporating advancements in natural language processing (NLP), such as word embeddings. Later, with the advent of transformer models like BERT, contextualized embeddings were integrated into topic modeling approaches, such as in Contextualized Topic Models (CTM).

The rise of LLMs, known for their ability to generate coherent and contextually relevant text, has opened up new possibilities for enhancing topic modeling. One notable development is the use of prompt engineering, exemplified by recent approaches like TopicGPT \cite{pham-etal-2024-topicgpt}. This method involves prompting an LLM to generate topics by processing each document individually. While this approach can produce high-quality topics, it incurs substantial computational costs due to the need to process prompts for every document, which can be particularly demanding with large corpora and complex LLMs. Additionally, prompting methods often lack a systematic framework for incorporating information such as authors and labels into the topic modeling process.

FANToM offers an alternative approach by integrating LLMs and topic modeling in a different way. It continues to use a VAE framework, known for its efficiency in topic modeling, while employing a smaller, task-specific LLM for labeling tasks. The VAE handles the core topic modeling process, effectively capturing the latent structure of the data and providing compact and interpretable topic representations. For labeling, FANToM utilizes a smaller, specialized LLM rather than a large-scale general-purpose model, which reduces computational demands while maintaining high accuracy. By aligning the outputs of the smaller LLM with the topics generated by the VAE, FANToM achieves a more coherent and adaptable solution that can be tailored to various datasets and applications.

Overall, FANToM opens up new possibilities for leveraging large-scale LLMs to enhance traditional topic modeling frameworks, leading to improved topic generation and alignment.

\setlength{\tabcolsep}{3pt}
\begin{table*}[ht]
    \centering
    \begin{tabular}{c|c|l|l|l|l|l|l|l|l|l|l|l|l}
    \hline \hline 
        & Models & \multicolumn{3}{c|}{20NG} & \multicolumn{3}{c|}{AGN} & \multicolumn{3}{c|}{DB-14} & \multicolumn{3}{c}{arxiv} \\
        & & TC & TD & TQ & TC & TD & TQ & TC & TD & TQ & TC & TD & TQ\\
        \hline
        \multicolumn{13}{c}{\textbf{50 Topics}} \\
        \hline
         \multirow{5}{*}{\rotatebox{90}{baseline}} 
         & SCHOLAR & $0.410$ & $0.57$ & $0.234$ & $0.372$ & $0.44$ & $0.164$ & $0.629$ & $0.68$ & $0.428$ & $0.364$ & $0.46$ & $0.168$\\
         & DVAE    & $0.354$ & $0.59$ & $0.209$ & $0.393$ & $0.46$ & $0.181$ & $0.658$ & $0.66$ & $0.434$ & $0.360$ & $0.48$ & $0.173$\\
         & ETM     & $0.418$ & $0.56$ & $0.234$ & $0.326$ & $0.43$ & $0.140$ & $0.630$ & $0.67$ & $0.422$ & $0.281$ & $0.45$ & $0.126$\\
         & CTM     & $0.412$ & $0.58$ & $0.239$ & $0.371$ & $0.44$ & $0.163$ & $0.632$ & $0.69$ & $0.436$ & $0.348$ & $0.47$ & $0.164$\\
         & ECRTM   & $0.465$ & $0.60$ & $0.279$ & $0.389$ & $0.45$ & $0.175$ & $0.670$ & $0.71$ & $0.475$ & $0.390$ & $0.49$ & $0.191$\\
         \hline
         \multirow{5}{*}{\rotatebox{90}{FANToM (ours)}} 
         & SCHOLAR & $0.415$ & $0.62$ & $0.257$ & $0.397$ & $0.48$ & $0.191$ & $0.608$ & $0.72$ & $0.438$ & $0.375$ & $0.50$ & $0.188$\\
         & DVAE    & $0.424$ & $0.61$ & $0.258$ & $0.409$ & $0.50$ & $0.205$ & $0.611$ & $0.71$ & $0.434$ & $0.336$ & $0.51$ & $0.171$\\
         & ETM     & $0.410$ & $0.60$ & $0.246$ & $0.391$ & $0.49$ & $0.191$ & $0.639$ & $0.70$ & $0.447$ & $0.310$ & $0.49$ & $0.151$\\
         & CTM     & $0.436$ & $0.64$ & $0.279$ & $0.395$ & $0.47$ & $0.186$ & $0.622$ & $0.69$ & $0.429$ & $0.342$ & $0.52$ & $0.178$\\
         & ECRTM   & $0.440$ & $0.65$ & $0.286$ & $0.400$ & $0.50$ & $0.200$ & $0.645$ & $0.73$ & $0.471$ & $0.395$ & $0.53$ & $0.209$\\
    \hline
    \multicolumn{13}{c}{\textbf{200 Topics}} \\
    \hline
         \multirow{5}{*}{\rotatebox{90}{baseline}} 
         & SCHOLAR & $0.362$ & $0.47$ & $0.170$ & $0.354$ & $0.38$ & $0.134$ & $0.581$ & $0.57$ & $0.331$ & $0.342$ & $0.41$ & $0.140$\\
         & DVAE    & $0.376$ & $0.49$ & $0.184$ & $0.374$ & $0.39$ & $0.146$ & $0.602$ & $0.58$ & $0.349$ & $0.335$ & $0.43$ & $0.144$\\
         & ETM     & $0.368$ & $0.48$ & $0.177$ & $0.361$ & $0.37$ & $0.134$ & $0.590$ & $0.59$ & $0.348$ & $0.328$ & $0.40$ & $0.131$\\
         & CTM     & $0.384$ & $0.50$ & $0.192$ & $0.368$ & $0.40$ & $0.147$ & $0.595$ & $0.60$ & $0.357$ & $0.340$ & $0.42$ & $0.143$\\
         & ECRTM   & $0.397$ & $0.52$ & $0.206$ & $0.370$ & $0.41$ & $0.151$ & $0.612$ & $0.61$ & $0.373$ & $0.355$ & $0.44$ & $0.156$\\
         \hline
         \multirow{5}{*}{\rotatebox{90}{FANToM (ours)}} 
         & SCHOLAR & $0.380$ & $0.54$ & $0.205$ & $0.362$ & $0.40$ & $0.145$ & $0.615$ & $0.62$ & $0.381$ & $0.355$ & $0.46$ & $0.163$\\
         & DVAE    & $0.387$ & $0.53$ & $0.205$ & $0.381$ & $0.42$ & $0.160$ & $0.617$ & $0.61$ & $0.376$ & $0.360$ & $0.48$ & $0.173$\\
         & ETM     & $0.379$ & $0.52$ & $0.197$ & $0.385$ & $0.41$ & $0.158$ & $0.620$ & $0.60$ & $0.372$ & $0.365$ & $0.45$ & $0.164$\\
         & CTM     & $0.399$ & $0.55$ & $0.220$ & $0.380$ & $0.39$ & $0.148$ & $0.624$ & $0.63$ & $0.393$ & $0.369$ & $0.49$ & $0.181$\\
         & ECRTM   & $0.404$ & $0.56$ & $0.226$ & $0.386$ & $0.43$ & $0.166$ & $0.630$ & $0.64$ & $0.403$ & $0.371$ & $0.50$ & $0.185$\\
         \hline
\end{tabular}
\caption{Topic Coherence (TC), Topic Diversity (TD), and Topic Quality (TQ) for baseline and FANToM models across datasets for 50 and 200 topics in semi-supervised setting. In general, FANToM outperforms baseline models across both 50 and 200 topics.}
\label{tab:semi-sup}
\end{table*}

\setlength{\tabcolsep}{13pt}
\begin{table*}[t]
    \centering
    \begin{tabular}{c|ll}
        \hline \hline
         Index/Label & Aspect&Topics/Authors  \\
         \hline
          2 & \textit{topic} & graphs, chromatic, graph, bipartite, isomorphic\\
           & \textit{authors} & Kráľ\_D, Seymour\_P, Alon\_N, Wood\_DR, Koolen\_JH\\
           \hline
           4 & \textit{topic} & exoplanet, orbiting, planet, kepler, comet\\
           & \textit{authors} & Ford Eric B., Agol E., Wright J.T., Pepe F., Henning T.\\
         \hline \hline
          math.CO & \textit{topic} & conjectured, automorphism, matroids, poset, extremal\\
           & \textit{authors} & Seymour\_P, Kráľ\_D, Rautenbach\_D, Klavžar\_S, Li\_X\\
           \hline
           astro-ph.EP & \textit{topic} & stellar, atmospheric, exoplanet, atmosphere, planet, orbiting\\
           & \textit{authors} & Henning T., Lagrange A. -M, Pepe F., Desidera S., Ford Eric B.\\
           \hline
           
    \end{tabular}
    \caption{Associations between Topics and Authors using FANToM(A) (top) and FANToM (bottom) approaches. FANToM builds meaningful correspondence between labels, topics, and authors.}
    \label{tab:authors}
\end{table*}

\begin{table*}[ht]
    \centering
    \begin{tabular}{c|c|c|c|c}
    \hline \hline 
        \multirow{2}{4em}{Datasets} & \multicolumn{2}{c|}{\textbf{\textit{ours (FANToM)}}} & \multicolumn{2}{c}{\textbf{\textit{baselines}}} \\
         \cline{2-5}
          & ETM & DVAE & ETM & DVAE \\
         \hline
         20NG & $119.29 \pm 1.34$&  $119.36 \pm 1.46$&  $115.84 \pm 1.76$& $119.46 \pm 1.36$\\
         AGN &   $157.78 \pm 2.02$&  $147.48 \pm 2.47$&  $148.05 \pm 3.06$&  $143.25 \pm 3.54$\\
         DB-14 & $155.27 \pm 4.03$&  $150.84 \pm 2.05$&  $147.16 \pm 2.99$&  $144.16 \pm 2.79$\\
         arxiv & $620.48 \pm 4.05$&  $605.06 \pm 5.88$&  $602.99 \pm 4.59$&  $595.23 \pm 6.82$\\
         \hline \hline
         arxiv$^{*}$ & $601.23 \pm 1.28$&  $590.71 \pm 8.53$&  -&  -\\
         arxiv$^{**}$ & $656.46 \pm 0.69$&  $601.85 \pm 9.88$&  -&  -\\
         
    \hline
    \end{tabular}
    \caption{This table presents a comparison of the runtime for the proposed FANToM model against baseline models. The upper part of the table displays the runtime values for FANToM(L) when only labels are incorporated, while the bottom part shows the runtime values for FANToM(A) (*) and FANToM (**) respectively. All runtimes are measured in seconds. The results indicate that incorporating the proposed methods does not have a significant impact on the runtime of the models.}
    \label{tab:runtime}
\end{table*}

\begin{table*}[ht]
    \centering
    \begin{tabular}{c|c|c}
    \hline \hline
    Datasets & \multicolumn{2}{c}{FANToM(L)} \\
    & \%MacroF1 & \%MicroF1  \\
    \hline
        20NG &  $0.523 \pm 0.02$ & $0.541 \pm 0.01$\\
        AGN &  $0.659 \pm 0.03$ & $0.655 \pm 0.03$\\
        DB-14 & $0.511 \pm 0.00$ & $0.532 \pm 0.01$\\
        arxiv & $0.707 \pm 0.03$ & $0.725 \pm 0.04$\\
    \hline
    \end{tabular}
    \caption{F1 scores of the FANToM(L) model validates the alignment of topics with labels.}
    \label{tab:classification}
\end{table*}

\setlength{\tabcolsep}{10pt}
\begin{table*}[ht!]
    \centering
    \begin{tabular}{c|c|c}
    \hline \hline
    Datasets & \multicolumn{2}{c}{FANToM(L)} \\
    & Top-3 & Top-5  \\
    \hline
        20NG &  $0.696 \pm 0.02$ & $0.798 \pm 0.01$\\
        AGN &  $0.892 \pm 0.02$ & $1.000 \pm 0.00$\\
        DB-14 & $0.726 \pm 0.01$ & $0.832 \pm 0.02$\\
        arxiv & $0.852 \pm 0.02$ & $0.918 \pm 0.01$\\
    \hline
    \end{tabular}
    \caption{Top-3 and Top-5 accuracy scores of the FANToM(L) model validates the alignment of topics with labels.}
    \label{tab:classification-acc}
\end{table*}

\begin{figure*}
    \centering
    \includegraphics[scale=0.6]{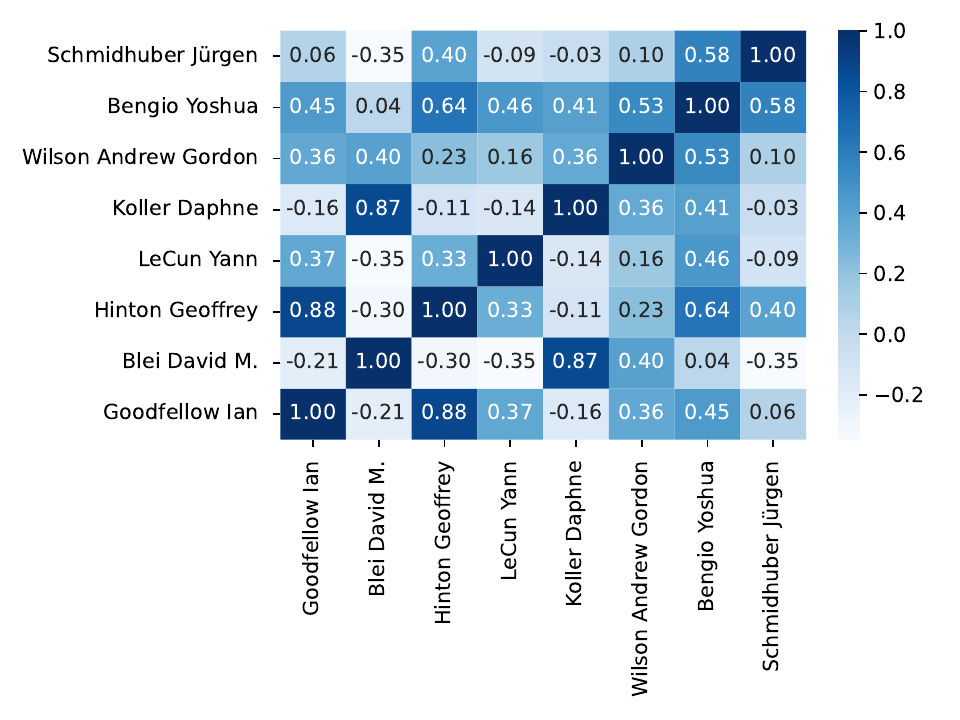}
    \caption{Cosine Similarity Matrix illustrating the relationships between prominent ML authors, determined by analyzing their associated topic vectors through the FANToM(A) model. High similarity scores indicate shared research areas among authors.}
    \label{fig:similarity}
\end{figure*}

\setlength{\tabcolsep}{8pt}
\begin{table*}[ht]
    \centering
    \begin{tabular}{c|c|c|c|c|c}
    \hline \hline 
        &Models & 20NG & AGN & DB-14 & arxiv \\
         \hline
         \multirow{5}{*}{\rotatebox{90}{baseline}} & L-LDA & $0.198 \pm 0.00$ & $0.336 \pm 0.00$ & $0.551 \pm 0.00$ & $0.269 \pm 0.02$\\
         & SCHOLAR & $0.410 \pm 0.01$ & $0.372 \pm 0.06$ & $0.629 \pm 0.02 \bullet$ & $0.364 \pm 0.03$\\
         & DVAE & $0.354 \pm 0.02$ & $0.393 \pm 0.07$ & $\textbf{0.658} \pm \textbf{0.01} \bullet$ & $0.360 \pm 0.03 \bullet$\\
         & ETM & $0.418 \pm 0.03 \bullet$ & $0.326 \pm 0.02$ & $0.630 \pm 0.01$ & $0.281 \pm 0.01$\\
         & CTM & $0.412 \pm 0.01$ & $0.371 \pm 0.01$ & $0.632 \pm 0.01 \bullet$ & $0.348 \pm 0.02 \bullet$\\
         \hline
         \multirow{4}{*}{\rotatebox{90}{ours}} & FANToM-SCHOLAR & $0.415 \pm 0.02 \bullet$ & $0.397 \pm 0.04 \bullet$ & $0.608 \pm 0.02$ & $\textbf{0.375} \pm \textbf{0.01} \bullet$\\
         & FANToM-DVAE & $0.424 \pm 0.03 \bullet$ & $\textbf{0.409} \pm \textbf{0.05} \bullet$ & $0.611 \pm 0.01$ & $0.336 \pm 0.02$\\
         & FANToM-ETM & $0.410 \pm 0.01$ & $0.391 \pm 0.02 \bullet$ & $0.639 \pm 0.02 \bullet$ & $0.310 \pm 0.02 \bullet$\\
         & FANToM-CTM & $\textbf{0.436} \pm \textbf{0.01} \bullet$ & $0.395 \pm 0.02 \bullet$ & $0.622 \pm 0.02$ & $0.342 \pm 0.01$\\
    \hline
    \end{tabular}
    \caption{Comparison of the proposed FANToM(L) (bottom) against the baselines (top) using Topic Coherence metrics. The best results across datasets are highlighted in bold, and best results across corresponding models between baselines and FANToM(L) are marked with $\bullet$. The proposed model generally outperforms the existing baselines.}
    \label{tab:benchmarking-coherence}
\end{table*}

\begin{table*}[ht]
    \centering
    \begin{tabular}{c|c|c|c|c|c}
    \hline \hline 
        &Models & 20NG & AGN & DB-14 & arxiv \\
         \hline
         \multirow{5}{*}{\rotatebox{90}{baseline}} & L-LDA & $0.97 \pm 0.00$ & $0.95 \pm 0.00$ & $0.93 \pm 0.00$ & $0.91 \pm 0.01$\\
         & SCHOLAR & $0.95 \pm 0.01$ & $1.00 \pm 0.00$ & $0.95 \pm 0.01$ & $0.94 \pm 0.01$\\
         & DVAE & $1.00 \pm 0.00$ & $1.00 \pm 0.00$ & $0.95 \pm 0.01$ & $0.94 \pm 0.01$\\
         & ETM & $0.95 \pm 0.01$ & $0.95 \pm 0.00$ & $0.96 \pm 0.01$ & $0.93 \pm 0.02$\\
         & CTM & $1.00 \pm 0.00$ & $1.00 \pm 0.00$ & $0.94 \pm 0.01$ & $0.93 \pm 0.02$\\
         \hline
         \multirow{4}{*}{\rotatebox{90}{ours}} & FANToM-SCHOLAR & $0.97 \pm 0.01$ & $1.00 \pm 0.00$ & $1.00 \pm 0.00$ & $1.00 \pm 0.00$\\
         & FANToM-DVAE & $1.00 \pm 0.00$ & $1.00 \pm 0.00$ & $1.00 \pm 0.00$ & $1.00 \pm 0.00$\\
         & FANToM-ETM & $0.97 \pm 0.01$ & $1.00 \pm 0.00$ & $0.99 \pm 0.00$ & $1.00 \pm 0.00$\\
         & FANToM-CTM & $1.00 \pm 0.00$ & $1.00 \pm 0.00$ & $1.00 \pm 0.00$ & $1.00 \pm 0.00$\\
    \hline
    \end{tabular}
    \caption{Comparison of the proposed FANToM(L) (bottom) against the baselines (top) using Topic Diversity metrics. The proposed model produces topics with higher diversity.}
    \label{tab:benchmarking-diversity}
\end{table*}
\end{document}